\documentclass[aps,pra,twocolumn,superscriptaddress]{revtex4-2}

\usepackage{amsmath}
\usepackage{amsfonts}
\usepackage{amssymb}
\usepackage{graphicx}
\usepackage{bm}
\usepackage{color}
\usepackage{multirow}
\usepackage{lineno}
\usepackage{dsfont}
\usepackage{siunitx} 
\usepackage{lipsum}
\usepackage{array}
\usepackage{newtxtext,newtxmath}
\usepackage{xcolor} 
\usepackage{cancel}
\usepackage{printlen}

\usepackage{verbatim}
\usepackage{mathtools}
\usepackage{empheq}

\usepackage[normalem]{ulem}  
\usepackage{cancel}

\definecolor{darkgreen}{rgb}{0.09, 0.55, 0.3}

\definecolor{darkred}{rgb}{0.8, 0.10, 0.1}

\newcommand{\bea}{\begin{eqnarray}}
\newcommand{\eea}{\end{eqnarray}}
\newcommand{\be}{\begin{equation}}
\newcommand{\ee}{\end{equation}}
\newcommand{\bes}{\begin{equation*}}
\newcommand{\ees}{\end{equation*}}
\newcommand{\bi}{\begin{itemize}}
\newcommand{\ei}{\end{itemize}}

\renewcommand{\vec}{\mathbf}
\DeclareSymbolFont{usualmathcal}{OMS}{cmsy}{m}{n}
\DeclareSymbolFontAlphabet{\mathcal}{usualmathcal}

\newcommand{\Id}{\ensuremath{\mathcal{I}}}
\newcolumntype{P}[1]{>{\centering\arraybackslash}p{#1}}
\newcommand{\pr}[1]{\ensuremath{\left[#1\right]}} 
\newcommand{\pc}[1]{\ensuremath{\left(#1\right)}}

\newcommand{\ket}[1]{\ensuremath{\left\vert#1\right\rangle}} 
\newcommand{\md}[1]{\ensuremath{\left\vert#1\right\vert}}

\definecolor{bluedarkRL}{rgb}{0,0.3,0.79}

\definecolor{burgundy}{rgb}{0.5, 0.0, 0.13}
\definecolor{denim}{rgb}{0.08, 0.38, 0.74}
\definecolor{midnightgreen}{rgb}{0.0, 0.29, 0.33}
\definecolor{sienna}{rgb}{0.53, 0.18, 0.09}
\definecolor{sacramentostategreen}{rgb}{0.0, 0.34, 0.25}

\newcommand{\mi}{\mathrm{i}}

\DeclareSIUnit\gauss{G}
\definecolor{Green}{rgb}{0,0.6,0.4}

\usepackage{xcolor}
\usepackage[%
  colorlinks=true,
  urlcolor=bluedarkRL,
  linkcolor=bluedarkRL,
  citecolor=bluedarkRL
]{hyperref}
\hypersetup{
colorlinks=true,
linkcolor=bluedarkRL,
citecolor=bluedarkRL,
urlcolor=black}

\newcommand{\alphaS}{399}
\newcommand{\alphaSerrstat}{30}
\newcommand{\alphaSerrsyst}{26}
\newcommand{\alphaV}{41}
\newcommand{\alphaVerrstat}{16}
\newcommand{\alphaVerrsyst}{3}

\newcommand{\alphaSNIST}{362}
\newcommand{\alphaSerrNIST}{20}
\newcommand{\alphaVNIST}{38}
\newcommand{\alphaVerrNIST}{15}

\begin{document}

\title{Determination of the ground state polarizability of $^{162}$Dy near $\SI{530}{\nano\meter}$
}

\author{Alexandre Journeaux}
\thanks{These authors contributed equally to this work.}
\affiliation{Laboratoire Kastler Brossel, Coll\`ege de France, CNRS, ENS-Universit\'e PSL, Sorbonne Universit\'e, 11 Place Marcelin Berthelot, 75005 Paris, France}
\author{Maxime Lecomte}
\thanks{These authors contributed equally to this work.}
\affiliation{Laboratoire Kastler Brossel, Coll\`ege de France, CNRS, ENS-Universit\'e PSL, Sorbonne Universit\'e, 11 Place Marcelin Berthelot, 75005 Paris, France}
\author{Julie Veschambre}
\affiliation{Laboratoire Kastler Brossel, Coll\`ege de France, CNRS, ENS-Universit\'e PSL, Sorbonne Universit\'e, 11 Place Marcelin Berthelot, 75005 Paris, France}
\author{Maxence Lepers}
\affiliation{Universit\'e Bourgogne Europe, CNRS, Laboratoire Interdisciplinaire Carnot de Bourgogne ICB UMR 6303, 21000 Dijon, France}
\author{Jean Dalibard}
\affiliation{Laboratoire Kastler Brossel, Coll\`ege de France, CNRS, ENS-Universit\'e PSL, Sorbonne Universit\'e, 11 Place Marcelin Berthelot, 75005 Paris, France}
\author{Raphael Lopes}
\email{raphael.lopes@lkb.ens.fr}
\affiliation{Laboratoire Kastler Brossel, Coll\`ege de France, CNRS, ENS-Universit\'e PSL, Sorbonne Universit\'e, 11 Place Marcelin Berthelot, 75005 Paris, France}

\begin{abstract}
Open-shell lanthanide atoms, and dysprosium in particular, combine a large ground-state angular momentum with dense electronic spectra, making their dynamical polarizability strongly dependent on wavelength and internal state, and therefore challenging to calculate and characterize experimentally.
This issue has become especially relevant with the recent development of single-atom trapping of dysprosium in optical-tweezer arrays, where precise knowledge of the polarizability is needed to design optimized trapping architectures. Here, we exploit the strong spin-dependent light shift near the $J'=J-1$ intercombination line at \SI{530.306}{\nano\meter} to determine the background scalar and vector polarizabilities of $^{162}$Dy in its ground state near this wavelength. Our measurements quantitatively agree with atomic-structure calculations and provide new insight into the contributions of nearby transitions in a spectral region relevant to emerging dysprosium tweezer platforms.
\end{abstract}

\maketitle

The frequency response of an atom to an oscillating electric field is characterized by its dynamical polarizability $\alpha(\omega)$, which can be written as a sum over dipole-allowed transitions weighted by dipole matrix elements \cite{Mitroy2010}. In the far-detuned regime, $\alpha(\omega)$ sets the light shift of an internal state and thus the optical dipole potential in a laser field \cite{Grimm2000a}. Accurate determination of the polarizability is therefore required both for quantitative modeling of optical traps and for precision metrology, where differential polarizabilities enter the evaluation of laser-induced and blackbody-radiation shifts in optical frequency standards \cite{Ludlow2015}.

Dysprosium combines a large magnetic-dipole moment ($\mu \simeq 10\,\mu_{\rm B}$) 
with a rich optical spectrum \cite{Lu2010,Chomaz2022,Norcia2021}. These properties have enabled studies of dipolar quantum gases \cite{Chomaz2022}, the emulation of artificial gauge fields in its large synthetic dimension \cite{Chalopin2020,Bouhiron2024}, as well as controlled light scattering in dense ensembles \cite{Hofer2025} 
and the trapping, imaging, and cooling of single Dy atoms in optical tweezer arrays at \SI{532}{\nano\meter} \cite{Bloch2023,Grun2024,Biagioni2025}.

In contrast with alkali-metal atoms, whose ground-state polarizabilities are predominantly scalar, lanthanides such as erbium and dysprosium exhibit sizable vector and tensor contributions, leading to spin- and polarization-dependent optical potentials~\cite{Li2017,Chalopin2018,Norcia2021, Tsyganok2019}. Predicting $\alpha(\omega)$ in the visible and near-infrared is particularly challenging for dysprosium because multiple transitions contribute appreciably, making theoretical values sensitive to the detailed atomic spectrum \cite{Li2017,Dzuba2011,Lepers2014}. While benchmark measurements exist at \SI{1064}{\nano\meter} \cite{Ravensbergen2018PRL} and near the \SI{626}{\nano\meter} intercombination line \cite{Kreyer2021}, recent measurements at \SI{532}{\nano\meter} report a ground-state scalar polarizability about a factor of two below theoretical expectations and comparable to its \SI{1064}{\nano\meter} value \cite{Bloch2024}. These observations motivate a dedicated calibration of the polarizability of Dy around 530\,nm, which is a key ingredient toward trapping and manipulating lanthanides in optical tweezers \cite{Bloch2023,Grun2024}.

\begin{figure}[t!]
    \centering
    \includegraphics[width=\columnwidth]{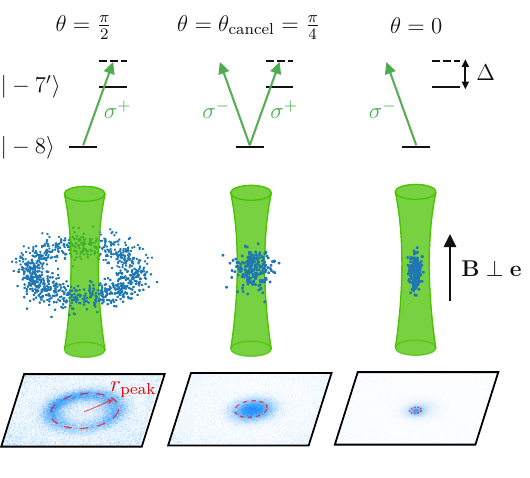}
\caption{Schematic illustration of the polarization dependence of the spin-dependent light shift near the $J\rightarrow J-1$ transition. The polarization is parameterized by the angle $\theta$, which determines the relative weights of the $\sigma^+$ and $\sigma^-$ components. In the configuration shown here, the $\sigma^+$ component gives a repulsive contribution through the coupling to the excited-state manifold, while the $\sigma^-$ component gives an overall attractive contribution dominated by the background polarizability. Top panels: corresponding optical couplings between the ground-state manifold ($J=8$) and the excited-state manifold ($J'=7$). For a polarization $\theta=\theta_{\rm cancel}$, equal to $\pi/4$ for the detuning used in this example, the light shift for the state $\ket{-8}$ vanishes. Bottom panels: qualitative effect of the resulting optical potential on the time-of-flight expansion of an atomic cloud. For $\theta=\pi/2$ (left), the light shift is repulsive and transfers potential energy into kinetic energy, leading to an increased expansion radius. At $\theta=\theta_{\rm cancel}$ (center), the light shift vanishes and the expansion is identical to the reference case without the SLS beam. For $\theta=0$ (right), the light shift is attractive and reduces the expansion. The peak radius $r_{\rm peak}$ extracted from the azimuthally integrated density profile is used experimentally to determine the cancellation condition.
}
    \label{figscheme}
\end{figure}

In this work, we exploit a narrow optical transition of $^{162}$Dy at $\lambda_{0}\simeq\SI{530.306}{\nano\meter}$ connecting the $J=8$ ground-state manifold to an excited state with $J'=J-1$~\cite{Wickliffe2000,Lecomte2025} with natural linewidth $\Gamma_0$. 
For detunings $\Delta$ satisfying $|\Delta|\gg \Gamma_{0}$ while remaining small compared with the detuning to neighboring optical transitions, the polarizability can be written as the sum of a resonant term associated with the $J\!\to\!J-1$ line and a background term arising from all other transitions and varying slowly with wavelength~\cite{Mitroy2010}. Their interplay yields a strongly spin- and polarization-dependent light shift: for appropriate polarizations some Zeeman sublevels are uncoupled from the excited manifold (dark states)~\cite{Lecomte2025, Journeaux2026}, and for specific detunings the total light shift of selected sublevels can vanish~\cite{Kao2017} (see Fig.~\ref{figscheme}). By locating these cancellation points as a function of detuning for the lowest Zeeman sublevel, we extract the background polarizability from a zero-crossing condition on the total polarizability. In contrast to the  determination of the light shift, this approach does not require an absolute calibration of the beam intensity or mode profile, which is often a major source of uncertainty in experimental platforms.




In the following, we consider the role of the vector and tensor parts of the polarizability in the determination of the light shift.
The optical dipole potential created by an off-resonant laser beam of intensity $I$ and unit polarization vector ${\bf e}$ is given by
\be
\hat V = - \frac{I}{2 \epsilon_0 c} \hat \alpha \ ,
\ee
where $\hat\alpha$ denotes the real part of the polarizability and defines the conservative light shift in the far-detuned, low-saturation regime. The polarizability operator can be written as
\begin{align}
\hat \alpha &= \alpha_s \hat \Id
- \mi \alpha_v ({\bf e}^* \times {\bf e}) \cdot \frac{{\bf \hat J}}{2J}  + \alpha_t \frac{1}{2J(2J-1)}  \nonumber\\
&\times \pr{ 3 \pc{({\bf e}^* \cdot {\bf \hat J})({\bf e} \cdot {\bf \hat J}) + ({\bf e} \cdot {\bf \hat J})({\bf e}^* \cdot {\bf \hat J})}  - 2\hat{\bf J}^{\,2} } \ ,
\end{align}
corresponding to a decomposition of the atomic polarizability into three components: a scalar part $\alpha_s$, a vector part $\alpha_v$, and a tensor part $\alpha_t$

We consider light near the $J=8 \rightarrow J'=7$ transition at $\lambda_0 = 530.3060(2)\,\mathrm{nm}$ \cite{Wickliffe2000,Note1}. \footnotetext{Unless otherwise stated, the number in parentheses stands for the statistical uncertainty corresponding to one standard deviation.} For detunings $\Delta=\omega-\omega_0$ small compared with the detuning to neighboring resonances, the polarizability can be decomposed as
\be
\alpha_{(s,v,t)}=\alpha^{\rm res}_{(s,v,t)}+\alpha^{\rm bg}_{(s,v,t)} ,
\ee
where $\alpha^{\rm bg}$ is a slowly varying contribution as a function of the wavelength.

\begin{figure}[t!]
    \centering
    \includegraphics[width=\columnwidth]{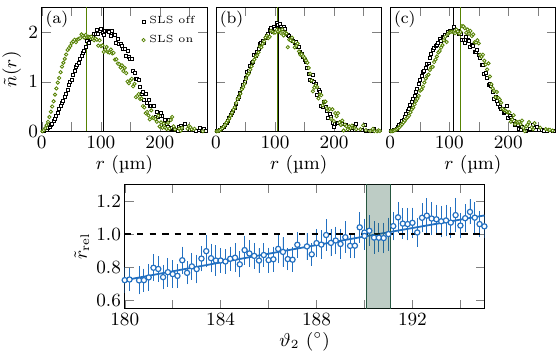}
\caption{Zero-crossing of the polarizability from time-of-flight expansion in the presence of the SLS beam at a detuning $\Delta/(2\pi) = \SI{815.9}{\giga\hertz}$ and a quarter-waveplate angle $\vartheta_4=160^\circ$. 
Top panels: examples of azimuthally integrated optical density $\tilde n(r)$ as a function of the distance from the cloud center, $r$, for three different polarizations of the spin-dependent light-shift (SLS) laser beam, with half-waveplate angles $\vartheta_2$ equal to (a) $180.2^\circ$, (b) $190^\circ$, and (c) $210.8^\circ$. Black curves correspond to the expansion without the SLS beam (reference taken independently for each value of $\vartheta_2$), while green curves show the expansion in the presence of the SLS beam. We realize (a) an attractive potential ($U_{\rm LS}<0$), (b) an almost vanishing potential ($U_{\rm LS}\approx 0$), and (c) a repulsive potential ($U_{\rm LS}>0$). Vertical dashed lines indicate the extracted peak radius $\tilde r_{\rm peak}$. 
Bottom panel: relative peak radius $\tilde r_{\rm rel}=\tilde r_{\rm peak}^{\rm(on)}/\tilde r_{\rm peak}^{\rm(off)}$ as a function of the angle of the half-waveplate, $\vartheta_{2}$. The dashed horizontal line at $\tilde r_{\rm rel.}=1$ marks the condition where the SLS beam does not modify the expansion, corresponding to a cancellation of the light shift. The grey shaded area indicates the uncertainty on the determination of $\vartheta_2^{\rm cancel}$, with its width corresponding to twice the standard deviation of the fitted value of $\vartheta_2^{\rm cancel}$.}
    \label{fig2}
\end{figure}

\begin{figure*}[t!]
    \centering
    \includegraphics[width=\textwidth]{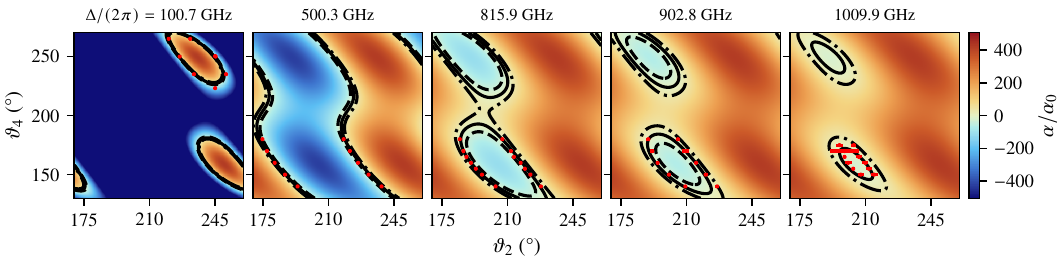}
    \caption{Polarizability cancellation for different detunings. Each panel shows the polarizability $\alpha$ (in units of $\alpha_0$) of the state \ket{-J} as a function of the half-waveplate angle $\vartheta_{2}$ and quarter-waveplate angle $\vartheta_{4}$, computed using the fitted background polarizabilities $\alpha^{\rm{bg}}_{st} = 399\,\alpha_0$ and
$\alpha^{\rm{bg}}_{v} = 41\,\alpha_0$. Since the polarizability is insensitive to the phase $\varphi$ and depends only on the degree of circular polarization $\theta$, continuous sets of combinations $(\vartheta_{2}, \vartheta_{4})$ yield the same value of $\alpha$.  
The panels correspond to different detunings $\Delta$ of the SLS beam, from left to right: $\Delta/(2\pi) = 100.7,\ 500.3,\ 815.9,\ 902.8,$ and $1009.9~\mathrm{GHz}$.  The Jones matrix $M$ (see main text) is the same for all panels. Red disks indicate the experimental zero crossings extracted from the expansion measurements, while solid black lines show the calculated zero-polarizability contours. Dashed and dash-dotted contours illustrate the expected shift of the zero-crossing lines when $\alpha^{\mathrm{bg}}_{\mathrm{st}}$ is varied by $+30\,\alpha_0$ and $-30\,\alpha_0$, respectively. For the largest detuning, the reduced sensitivity of the polarizability to polarization prevents the $-30\,\alpha_0$ contour from appearing.}
\label{fig3}
\end{figure*}

We first focus on the resonant contribution. For a sufficiently large bias magnetic field, couplings between different Zeeman states can be neglected and $\hat V$ is diagonal in the $\ket{J,m_J}$ basis. This secular approximation is valid when the Zeeman splitting between coupled Zeeman states is much larger than the off-diagonal light-shift couplings. In our experimental conditions, this requirement is already safely satisfied for $B\gtrsim\SI{150}{\milli\gauss}$, well below the field used here, \emph{i.e.} $B = 1.66~{\rm G}$. In the case of a beam propagating along the quantization axis, with $\mathbf{e}\perp\mathbf{B}$, the light shift of a state $\ket{J,m_J}$ then becomes
\be
\label{Ugeneral_reduced}
U_{\rm LS}
= \frac{D^2}{\Delta} \frac{I}{2\epsilon_0 c}
\Biggl[
\alpha_1
- \alpha_2 \cos(2\theta)\frac{m_J}{2J}
+ \alpha_3 \frac{J(J+1)-3m_J^2}{2J(2J-1)}
\Biggr] ,
\ee
where $\alpha_{(s,v,t)} = -(D^2/\Delta)\,\alpha_{(1,2,3)}$ with $(\alpha_1,\alpha_2,\alpha_3) = (5/17,-15/17,-5/17)$. Here $D^2 = d^2/\hbar = 3\pi\epsilon_0 (c/\omega_0)^3 \Gamma_0$, where $d$ is the ordinary dipole matrix element. The angles ($\theta$, $\varphi$) define the polarization, such that
\be
\mathbf e = \cos \theta\, {\bf e_{-}} + e^{-\mi \varphi} \sin \theta \, {\bf e_{+}} 
\ee
with ${\bf e_\pm}$ corresponding to the $\sigma^\pm$ polarizations. Equation~\eqref{Ugeneral_reduced} illustrates the polarization dependence of the light shift.  More generally, for a $J \to J-1$ transition, the coupling between ground and excited manifolds has a two-dimensional null space, since the ground-state manifold has dimension $2J+1$ while the excited-state manifold has dimension $2J-1$. As a result, there are always two dark states for arbitrary polarization. For $J=8$, pure $\sigma^-$ polarization ($\theta=0$) cancels the light shift for $m_J=-7,-8$.

We now include the background contribution. Restricting to atoms prepared in the lowest-energy Zeeman sublevel $\ket{J,-J}$, it is convenient to introduce the normalized light shift  $\tilde U_{\rm LS}~=~(4\epsilon_0 c\, U_{\rm LS})/I$. One obtains
\be
\tilde U_{\rm LS} = - \Bigl(2 \alpha_s^{\rm bg} - \alpha_t^{\rm bg} - \frac{15}{17}\frac{D^2}{\Delta}\Bigr) - \cos(2\theta) \Bigl(\alpha_v^{\rm bg} + \frac{15}{17}\frac{D^2}{\Delta}\Bigr) \ ,
\label{ULS_reduced}
\ee
which can be expressed as the sum of two competing contributions: one involving the scalar and tensor polarizabilities, $\tilde U_{\rm LS}^{(st)}(\Delta)$, and one arising from the vector polarizability, $\cos(2\theta)\,\tilde U_{\rm LS}^{(v)}(\Delta)$. For a given detuning $\Delta$ such that $\md{\tilde U_{\rm LS}^{(st)} / \tilde U_{\rm LS}^{(v)}} \le 1$, there exists an angle $\theta_{\rm cancel}$ for which the total light shift vanishes.
In the following, we use these zero-crossing conditions to determine the background polarizability of $^{162}$Dy near \SI{530}{\nano\meter}.

Experimentally, we probe the sign and cancellation of the spin-dependent light shift (SLS) by monitoring the time-of-flight (ToF) expansion of a nondegenerate ($T\simeq\SI{200}{\nano\kelvin}$) cloud of $^{162}$Dy with $N\simeq 10^{5}$ atoms~\cite{Lecomte2024}. The sample is prepared in the lowest Zeeman sublevel $\ket{J,-J}$ in the presence of a bias magnetic field $\vec B = B\,\vec e_z$ with $B=\SI{1.66}{\gauss}$. In the following, we simplify the notation and use $\ket{J,m_J}\equiv\ket{m_J}$ and $\ket{J',m'_J}\equiv\ket{m'_J}$. After release from the trap, the atoms expand for $t_{\rm ToF}=\SI{30}{\milli\second}$ while illuminated by a laser beam propagating along $\vec e_z$ (SLS beam, Fig.~\ref{figscheme}) with $1/e^{2}$ radius $w=\SI{63.6(21)}{\micro\meter}$. The initial in-situ cloud radius is typically smaller than \SI{10}{\micro\meter}, much smaller than the SLS beam waist, such that all atoms experience nearly the same optical potential at the beginning of the expansion. The SLS beam is detuned from the $\lambda_0=\SI{530.306}{\nano\meter}$ transition; the laser wavelength ranging from \SI{529.37}{\nano\meter} to \SI{530.21}{\nano\meter}. For each detuning, the laser frequency is stabilized using a separate iodine reference. The transition frequency is determined independently from the maximum radiation pressure observed in a ``push'' experiment~\cite{Lecomte2025}.

The polarization is controlled using a half-waveplate and a quarter-waveplate with fast-axis angles $\vartheta_{2}$ and $\vartheta_{4}$ relative to the input linear polarization. We model by a Jones matrix, $M$, the transformation of the polarization along
the optical path between these waveplates and the atoms.

To determine the cancellation angle $\theta_{\rm cancel}$ at which the light shift of $\ket{-J}$ vanishes, we compare the time-of-flight expansion in the presence of the SLS beam to a reference expansion without the SLS beam. We compute the azimuthally integrated density profile $\tilde n(r)~=~\int {\rm d}\phi\, r\, n(r,\phi)$, and define the peak radius $\tilde r_{\rm peak}$ as the radius at which the azimuthally integrated density profile reaches its maximum value (see Fig.~\ref{fig2}). For each setting $(\vartheta_{2},\vartheta_{4})$, we compute the ratio
\be
\tilde r_{\rm rel}=\tilde r_{\rm peak}^{\rm(on)}/\tilde r_{\rm peak}^{\rm(off)},
\ee
and identify $\theta=\theta_{\rm cancel}$ from the condition $\tilde r_{\rm rel}=1$, which signals the cancellation of the light shift in the regime where the density profile is not significantly distorted.

\begin{figure*}[t]
    \centering
    \includegraphics[width=\textwidth]{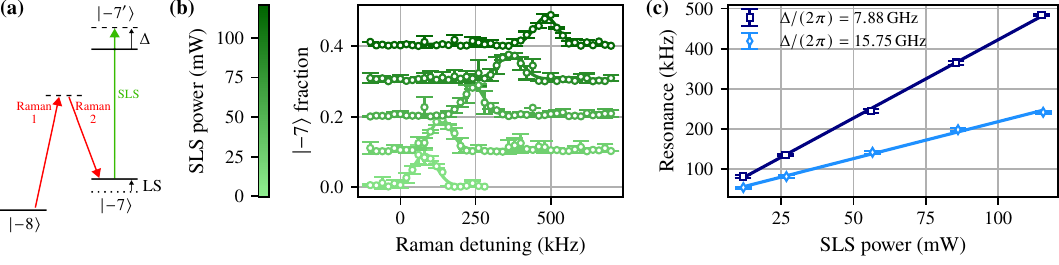}
    \caption{Calibration of the linewidth $\Gamma_0$. (a) Schematic representation of the spectroscopy method (see text). (b) Fraction of atoms transferred to the state $\ket{-7}$ by the Raman coupling, for an SLS detuning of $\Delta/(2\pi) = \SI{7.88}{\giga\hertz}$ and for different SLS powers, as a function of the two-photon detuning measured with respect to the Raman resonance in the absence of the SLS beam. The resonant condition depends on the intensity of the SLS beam. An offset is added to each curve for visibility. (c) Variation of the Raman resonance frequency with SLS power for two detunings. The linear dependence gives the SLS-induced light shift per unit power. Since, close to resonance, $U_{\rm LS}\propto I\,\Gamma_0/\Delta$, the slopes measured at different detunings provide a calibration of the linewidth $\Gamma_0$.}
    \label{fig:fig4}
\end{figure*}

For a given detuning $\Delta$, several waveplate settings yield the same value of $\theta_{\rm cancel}$.
This results in lines along which the polarizability vanishes, as shown in Fig.~\ref{fig3} (red disks). Repeating this procedure for multiple detunings, and assuming that the background polarizabilities remain constant over the explored \SI{1}{\tera\hertz} range~\footnote{This assumption is supported by atomic-structure calculations. At $\lambda=\SI{530.307}{\nano\meter}$, we obtain $\alpha_s=381.4\,\alpha_0$, $\alpha_v=39.9\,\alpha_0$, and $\alpha_t=14.9\,\alpha_0$, while at $\lambda=\SI{529.465}{\nano\meter}$, separated by approximately $\SI{0.9}{\tera\hertz}$, we obtain $\alpha_s=384.2\,\alpha_0$, $\alpha_v=39.1\,\alpha_0$, and $\alpha_t=15.0\,\alpha_0$. The corresponding relative variations are below $2.5\%$ over this range. In addition, the closest reported transitions in the NIST database, Ref.~\cite{Wickliffe2000} (see also \cite{Li2017}) are located at $\lambda=\SI{534.418}{\nano\meter}$ and $\lambda=\SI{523.771}{\nano\meter}$, well outside the explored frequency interval.}, we perform a global fit of all zero-crossing points to extract the Jones matrix $M$ as well as the background combinations
 $\alpha_v^{\rm bg}$ and $\alpha_{st}^{\rm bg}\equiv\alpha_s^{\rm bg}-\alpha_t^{\rm bg}/2$~\footnote{Because the SLS beam propagates along the quantization axis, $\alpha_s^{\rm bg}$ and $\alpha_t^{\rm bg}$ cannot be determined independently in our geometry.}. 
The experimental input to the global fit is the set of zero-crossing points $(\vartheta_{2,i},\vartheta_{4,i})$ measured at different detunings $\Delta_i$. For each point, the waveplate settings are converted into the polarization seen by the atoms using the waveplate matrices and the effective Jones matrix $M$. This polarization determines the angle $\theta_i$ entering the theoretical expression for the total polarizability, see Eq.~\eqref{ULS_reduced}. The fitted parameters are the parameters of $M$, together with the two background polarizability combinations $\alpha_{st}^{\rm bg}$ and $\alpha_v^{\rm bg}$.\\
The fit is performed by minimizing the residual total polarizability evaluated at all experimentally measured zero-crossing points. In the minimization, the residuals are weighted by the corresponding detuning $\Delta_i$ in order to compensate for the dominant $1/\Delta$ scaling of the resonant contribution and to give comparable weight to measurements taken at small and large detunings.
Once the optimal values of $M$, $\alpha_{st}^{\rm bg}$, and $\alpha_v^{\rm bg}$ are obtained, the polarizability maps shown in Fig.~\ref{fig3} are calculated from Eq.~\eqref{ULS_reduced} by scanning the waveplate angles over the full $(\vartheta_2,\vartheta_4)$ plane. The solid black curves correspond to the calculated contours for which the total polarizability vanishes,
and are in good agreement with the experimental data.

From the global fit we obtain, near $\lambda_0 = \SI{530.306}{\nano\meter}$, $\alpha_{st}^{\rm bg}/(\alpha_0 \Gamma_0) = \SI{426.7(69)}{\micro\second}$ and $\alpha_{v}^{\rm bg}/(\alpha_0 \Gamma_0) = \SI{44(17)}{\micro\second}$, where $\alpha_0=4\pi\epsilon_0 a_0^3$ is the atomic unit of polarizability. Determining the absolute values of the polarizabilities therefore only requires an accurate estimate of the transition linewidth $\Gamma_0$.

To obtain an independent experimental calibration of $\Gamma_0$, we measure the SLS-induced light shift of the state $\ket{-7}$ using a spectroscopy scheme akin to Autler--Townes spectroscopy~\cite{Autler1955,Zhang2013}.
Experimentally, we prepare a nondegenerate sample at a temperature of \SI{300}{\nano\kelvin} in an external magnetic field $\vec{B}=B\,\vec{e}_x$, with $B=\SI{2.28}{\gauss}$, where $\vec e_x$ is orthogonal to the propagation direction of the SLS beam. With this field orientation, the SLS beam is $\pi$-polarized, and the two dark states are then $\ket{\pm J}$. The SLS beam is applied at various intensities and detunings $\Delta$. We determine the light shift of the state $\ket{-7}$ by measuring the transition frequency $\ket{-8}\rightarrow\ket{-7}$ using a two-photon Raman coupling between these states, implemented with two co-propagating laser beams detuned from the optical transition at $\lambda=\SI{626.1}{\nano\meter}$ (see Fig.~\ref{fig:fig4}).

After the Raman pulse, a Stern--Gerlach time-of-flight sequence is used to determine the fraction of atoms transferred to $\ket{-7}$. Although the Raman beams induce an additional constant light shift on both states, differential measurements remain sensitive only to the contribution of the SLS beam. From the dependence of the Raman resonance frequency on the SLS intensity and detuning, and using the analogue of Eq.~(\ref{Ugeneral_reduced}) for $\pi$-polarized light, we extract the ordinary dipole matrix element $d$, and thus the linewidth $\Gamma_0$.

We obtain $\Gamma_0 /(2\pi) = 149(11)_{\rm stat}(10)_{\rm syst} \, \rm{kHz}$, where the systematic uncertainty arises from the calibration of the SLS beam waist, measured independently. Using this value of $\Gamma_0$, we obtain
\begin{align}
\alpha_{st}^{\rm bg} &= 
\alphaS(\alphaSerrstat)_{\rm stat}(\alphaSerrsyst)_{\rm syst}\,\alpha_0
\, , \nonumber \\
\alpha_{v}^{\rm bg}  &= 
\alphaV(\alphaVerrstat)_{\rm stat}(\alphaVerrsyst)_{\rm syst}\,\alpha_0
\, .
\end{align}
This result agrees within error bars with the value reported in the NIST database \cite{Wickliffe2000}, $\Gamma_0/(2\pi) = \SI{135(7)}{\kilo\hertz}$, the latter leading to background polarizabilities
\begin{align}
\alpha_{st}^{\rm bg} &= \alphaSNIST(\alphaSerrNIST)\,\alpha_0  \nonumber \\
\alpha_{v}^{\rm bg}  &= \alphaVNIST(\alphaVerrNIST)\,\alpha_0 \ .
\end{align}

We compare these results with theoretical values from our atomic-structure calculations. We use the sum formula over dipole-allowed transitions, with the same energies and transition dipole moments as in Ref.~\cite{Bloch2024}, except that we exclude the 530.306~nm transition from the sum. We also estimate the uncertainty on the calculated polarizabilities, by assuming that each line strength of the sum is computed with an uncertainty of 12.4~\%, (see Ref.~\cite{Lepers2025}~ Chapter~2). 
This value comes from our least-squares fit on Einstein coefficients \cite{Li2017}: it corresponds to the ratio between the standard deviation and the largest experimental Einstein coefficient included in the fit \cite{Wickliffe2000}.

We obtain the following results: $\alpha_{st}^{\rm bg, \ theory} = 374 (81) \ \alpha_0$ and $\alpha_{v}^{\rm bg, \ theory} = 40 (82) \ \alpha_0$, which are in excellent agreement with our above measurements. 
Within the \SI{529.37}{}-\SI{530.21}{\nano\meter} interval probed here, we do not observe evidence for features that would account for the discrepancy reported in Ref.~\cite{Bloch2024} at \SI{532.208}{\nano\meter}. Its origin therefore remains unclear.

To conclude, we have measured the scalar plus tensor component and vector component of the dynamical polarizability of $^{162}$Dy in the vicinity of the \SI{530.306}{\nano\meter} intercombination line. Our approach exploits polarization-dependent cancellations of the light shift at various detunings from the $J' = J-1$ transition. By tracking these zero crossings, we extract the background polarizability without relying on an absolute intensity calibration, making the method both robust and experimentally straightforward. 
A precise determination of the background polarizability near \SI{530}{\nano\meter} is directly relevant for experiments employing optical tweezers~\cite{Biagioni2025}, where vector and tensor contributions generate sizable differential light shifts between different electronic levels. More generally, engineered differential light shifts constitute a resource for advanced cooling and evaporation protocols in optical dipole traps which rely sensitively on accurate polarizability values~\cite{Lopes2021}. The improved precision reported here supports the implementation of such cooling strategies in cold dysprosium gases lanthanide systems.




\vspace{0.5cm}

\textit{Acknowledgments.} We thank Igor Ferrier-Barbut for fruitful discussions and for carefully reading the manuscript. We are also grateful to the members of the Bose--Einstein condensate team at LKB for helpful discussions. Calculations have been performed using HPC resources from DNUM CCUB (Centre de Calcul de l’Université de Bourgogne). This research was funded, in part, by the Agence Nationale de la Recherche (ANR) under projects ANR-20-CE30-0024 and ANR-24-CE30-7961. This work was also supported by the Région Île-de-France within the framework of DIM QuanTiP. For the purpose of open access, the author has applied a CC-BY public copyright licence to any Author Accepted Manuscript (AAM) version arising from this submission. 

\textit{Data availability}: The data that support the findings of this article are openly available  \cite{DataRepository}.

\bibliography{bib-final_AJ_revised.bib}

\end{document}